\documentstyle{article}
\begin{document}
\LARGE
\begin{center}\bf The Imaginary Time in the Tunneling Process

\vspace*{0.7in} \normalsize \large \rm

Wu Zhong Chao

Center for Astrophysics

University of Science and Technology of China

Hefei, Anhui, China

\vspace*{0.3in}

(July 25, 1995)

\vspace*{0.5in} \large \bf Abstract
\end{center}
\vspace*{.2in} \rm \normalsize By using techniques developed in
quantum cosmology, it is found that a tunneling particle spends
purely imaginary time on a barrier region. The \it imaginary \rm
time is associated with the stochastic acausal behaviour of a state,
while the \it real \rm time is associated with the deterministic
causal evolution of a state. For the tunneling case the nonzero
imaginary time is associated with the transmission rate of the
tunneling process, which is related to the thickness of the barrier.
The physical meaning of the zero real time is that the particle
instantly jumps from one side of the barrier to the other regardless
of the thickness. This leads to the illusion that tunneling
particles could actually travel faster than light. The results of
recent experiments in quantum optics concerning tunneling time can
be thought of as the first experimental confirmation of the
existence of imaginary time. Relativity is not violated.

\vspace*{0.45in}


\pagebreak

In special relativity it is postulated that light travels in a
vacuum at a constant speed (we set the vacuum speed of light $c$ to
be $1$) with respect to any inertial frame and nothing can travel
faster than light. This postulate leads to causality in relativity.
In general relativity, one can always formulate physical laws with
respect to local inertial frames at any points of a spacetime. Since
in general spacetime is curved, the statements on the speed of light
should refer to a local frame only.

In 1932 MacColl [1] argued that a particle tunnels through a
barrier without any appreciable delay. Later in 1955, Wigner and
Eisenbud [2] claimed that under some circumstances, tunneling
particles could travel faster than 1. Recently, experiments in
quantum optics by Chiao, Kwiat and Steinberg [3] seem to confirm
this. The aim of this paper is to clarify this
issue. I want to show that the propagation of light with a
speed greater than 1 is just an illusion: it is the effect of the
imaginary time spending. Therefore, both relativity and
causality remain intact.

Let us begin with the case of light propagation in a transparent
dispersive media.  This problem was quite thoroughly discussed by
Sommerfeld and Brillouin [4]. The
phase velocity is equal to $\omega/k$, where $w$ and $k$ are
frequency and wave number respectively. The group velocity is
defined as $d\omega/dk$. It was generally argued that the
information propagation should be carried at the group velocity.
It was implicitly assumed that the group velocity will never
exceed $1$ in reality, even theoretically it is quite possible
that the group velocity can exceed $1$, and at certain
frequencies can become infinity, or even negative. However,
recent experiments of superluminal propagation show that the
group velocity can exceed the speed of light [5]. Therefore, to
reexamine the speed of information propagation is inevitable.
Another relevant velocity is the so-called energy velocity; it is
defined by the ratio of time-averaged Poynting vector and the
time-averaged energy density. For transparent and dispersionless
media all these velocities are identical.

One can derive the Kramers-Kronig relations of the complex linear
susceptibility $\chi$ solely from the assumption of linearity and
causality of wave propagation in the media

\begin{equation}
Re \chi (\omega) = \frac{2}{\pi}P \int_0^\infty
\frac{\omega^\prime Im \chi (\omega\prime)}{\omega^{\prime 2} -
\omega^2}d\omega^\prime,
\end{equation}

\begin{equation}
Im \chi (\omega) = - \frac{2\omega}{\pi} P \int_0^\infty \frac{Re
\chi (\omega^\prime)}{\omega^{\prime 2} - \omega^2}
d\omega^\prime,
\end{equation}
where $P$ denotes the principal value. It is expected that all
linear media should satisfy the quite universal Kramers-Kronig
relations. If one elaborates the problem on wave propagation by
using the Shannon theory of information, he will find that the
information velocity will never exceed $1$, even for the case
with group velocity $v_g > 1, \rightarrow \infty$ or $< 0$. It
means that causality is not violated. One can expect that
causality will remain valid even for nonlinear media which does
not
respect the Kramers-Kronig relations.

The paradox, raised by the experiments performed by Chiao, Kwiat
and Steinberg [3] that a tunneling particle seemingly can travel
faster than $1$, is offered a resolution by these authors as
follows
[6]: In a typical experiment, the whereabouts of the photon,
detected only once, is best predicted by the location of the
peak of the wave packet. The wave packet of the tunneling photon
gets reshaped, and
the peak of the tunneling photon precedes that of a photon
traveling unimpeded at the speed of light. They believe that as
far as the wavefronts of these photons are concerned, at no point
does the tunneling-photon wave packet travel faster than the
free-traveling photon.

However, in the opinion of this author, this paradox cannot
easily be dispelled this way in the framework of quantum optics.
The
issue is far more fundamental than it looks. I believe that the
issue is
just a manifestation of the very nature of time concept.

I would like to present a very simple calculation of time spent
by a tunneling particle in the barrier region. In nonrelativistic
quantum mechanics, the Schroedinger equation of motion in one
dimension for a particle with mass $m$ takes the form:
\begin{equation}
i \frac{\partial \Psi}{\partial t} = -
\left(\frac{\partial^2}{2m\partial x^2} - U(x)\right ) \Psi,
\end{equation}
where $U$ is the potential of the field and we have set
$\hbar = 1$. Since the potential is time-independent, one can
solve the equation by using the complete set of the
stationary states which satisfy
\begin{equation}
\left(\frac{\partial^2\psi}{2m\partial x^2}\right ) + [E - U(x)]
\psi = 0,
\end{equation}
where $E$ is the energy of the stationary states.

In classical mechanics a particle with energy lower than the
height of a potential barrier is forbidden to overpass it.
In quantum mechanics, it turns out that under this circumstance
the particle can tunnel through the barrier, instead. By using
the WKB method, one can get the transmission coefficient for the
barrier
\begin{equation}
D \approx \exp \left[- 2 \int_a^b\sqrt{2m(U(x) - E)}dx\right ],
\end{equation}
where the integral is taken for the barrier region $[a, b]$,
where $U (x) \ge E$, $a, b$ are the turning points satisfying
$U(a) = U(b) = E$.

If one turns to the imaginary time regime by setting $\tau = it$,
then the potential and energy would reverse their signs, and the
particle motion can be described as a bounce solution in the
potential well. Therefore, the exponent in the transmission
coefficient formula can be rewritten as the negative of the
action for the classical bounce of the particle in the
imaginary time
\begin{equation}
A = \oint pdx,
\end{equation}
where $p$ is the momentum of the bounce solution. It was based on
this observation that Coleman developed the instanton theory.

The instanton theory has been widely accepted by the particle
physics
community. In particular, it has fundamental influence on
Euclidean quantum gravity and the no-boundary universe. Despite
this, people may still consider the above argument merely as a
calculation trick in quantum mechanics or quantum field theory.
It is my opinion that the current experimental results of quantum
optics is a clear confirmation of the physical existence of
imaginary time.

The most convenient way to investigate the time problem is the
Feynmann path integral approach, as in quantum cosmology [7].
Since
in quantum gravity, spacetime itself should be quantized,
therefore the time coordinate does not appear explicitly in the
path integral.
Indeed, a history from an initial closed 3-surface to a final
3-surface is represented by a 4-manifold sandwiched between them,
and the time lapse is somehow implied by the manifold. One
obtains the wave function by summation of all these histories,
and the wave function takes the superposition form of wave
packets

\begin{equation}
\Psi \approx \exp iS =\exp [i(S_r + iS_i)],
\end{equation}
where $S_r$ and $S_i$ are the real and imaginary parts of the
phase.

It is only after one obtains the wave function, that the time
concept will
appear explicitly. One can interpret the oscillatory
components associated with $S_r$ of the wave function as
classical evolutions in real time and the exponential components
associated with $S_i$ as classical evolutions in imaginary time.

For a system with only one degree of freedom , these two
behaviors are mutually exclusive in the configuration space. In
the higher dimensional case, in general they are coupled, and
these two classical trajectories
are mutually perpendicular in the configuration space. One can
experience the trajectory in real time as the deterministic and
causal evolution. A trajectory in imaginary time assigns
probability $\infty$ $\exp -2S_i$ to the ensemble of real time
trajectories it intersects.

In a closed universe, there does not exist an external time
coordinate. What one obtains is the intrinsic time of the
universe. For the Hawking massive scalar model [8], time is
imaginary
in the Euclidean regime and becomes real in the Lorentzian
region. In quantum mechanics, in general, the
external time is given, and the time derived from the wave
function can be identical to the intrinsic time
of the particle. Since we are dealing with the issue of how
much time a tunneling particle spends on the barrier region, the
particle time itself should be quantized, as in quantum
cosmology, and one has to adopt the path integral approach. The
action from initial position $x_i$ to final position $x_o$ is
written as
\begin{equation}
I = \int^{x_o}_{x_i}\left [\frac{m\dot{x}^2}{2} - U(x)\right ]
dt,
\end{equation}
where the dot denotes time derivative for the history.

We assume that the potential approached zero on the two far sides
of the barrier
\begin{equation}
\lim_{x \rightarrow \pm \infty} U(x) = 0,
\end{equation}

The wave function for a particle with energy $E$ initiated at $x
= x_i$ or $x = x_f$ becomes
\begin{equation}
\psi (x) = \int d[x] \exp (- \bar{I}(x)),
\end{equation}
the path integral is summed over all trajectories with a fixed
momentum
corresponding to the energy $E$ at $x= x_i$ or $x = x_o$. The
original path integral is divergent in real time, therefore we
have to evaluate it in the Euclidean regime, i.e., we have made
the Wick rotation by defining the imaginary time $\tau = it$ and
$\bar{I}\equiv -iI$ is the Euclidean action.

The path integral over trajectories with a fixed momentum at the
right far side of the barrier $x_o$ $(x_o > b)$ represents the
stationary state of a particle propagating to the right hand
side.
The particle from the left is reflected by and penetrated through
the barrier.
This boundary condition is similar to that in cosmological
models. For example, in the
Hawking model [8], the derivative of the universe scale with
respect
to the imaginary time is set to be 1 at the south pole of the
Euclidean sector of the spacetime manifold.

The main contribution to the path integral comes from the
classical solutions. For the classical solutions, the action
becomes
\begin{equation}
I = \int pdx,
\end{equation}
where $p$ is the canonical momentum. The classical solutions with
real action in real time dominate the wave function outside the
barrier, while the bounce solution with imaginary action in
imaginary time controls the quantum behaviour within the barrier.
Substituting (11) into the path integral, one gets
\begin{equation}
\psi = -\sqrt{ \frac{m}{p}}\exp \left [ -\left |\int^b_a
pdx\right| + i \int^x_b pdx + \frac{1}{4} i\pi \right ], \;\;\;\;
x> b
\end{equation}
\begin{equation}
\psi = \sqrt{\frac{m}{p}}\exp \left [ - \left|\int^b_a pdx \right
| + \left| \int^x_b pdx\right| \right], \;\;\;\; a <x<b
\end{equation}
\begin{equation}
\psi = \sqrt{\frac{m}{p}} \exp \left [ i \int^x_a pdx +
\frac{1}{4} i\pi \right ] + \sqrt{\frac{m}{p}} \exp \left [ -i
\int^x_a pdx - \frac{1}{4} i\pi \right ], \;\;\;\; x<a
\end{equation}
where we have worked the higher order quantum correction by
solving the equation (4) through substitution. For simplicity,
we have used the approximation that the transmission coefficient
$D \ll 1$, and the chosen normalization corresponds to
the unit probability current density in the incident wave from
the left side. For the general case, the solution can be obtained
as a superposition of multiple reflections and transmissions at
the boundaries of the barrier. It is helpful to note that the
particle traveling on the barrier region takes no real time, as
we shall show below.

The truncated Schroedinger equation (4) can be thought of as the
counterpart of the Wheeler-DeWitt equation in
quantum cosmology, in which the time coordinate does not appear
explicitly.

To recover the particle time, one can identify
\begin{equation}
p = \frac{\partial S}{\partial x},
\end{equation}
and, as expected, the time spent outside (inside) the barrier is
real
(imaginary), corresponding to the classical (bounce) solution.
The imaginary time spent on the barrier region is
\begin{equation}
t = \int_b^a \frac{mdx}{p}.
\end{equation}
Since both the momentum and time are imaginary within the
barrier,
then the wave function decays exponentially and it leads to the
transmission rate (5).

I believe that this is the simplest way to derive the
time the tunneling particle spent on the barrier region. The
result is consistent with the Sokolovski and Connor's calculation
[9]. Baz' and Rybachenko [10]
have proposed the use of the Larmor precession as a clock to
measure the time it takes a particle to traverse a barrier. If
the particle carries spin $1/2$ and a small magnetic field
pointing in a direction perpendicular to the spin is confined to
the barrier, then the magnetic field does not actually perform a
Larmor precession. Its only effect is to align the spin with the
field. The reason is that in the imaginary time, the kinetic
energy is split by the Zeeman effect contribution $\pm w_L/2$,
where $w_L$ is the Larmor frequency. The energy difference causes
the difference of transmission rates. It results
experimentally in a spin components $w_L |t|/2$ along the
magnetic field. This method has been used to obtain highly
polarized electrons from metals coated with a thin film of a
ferromagnetic semiconductor.

In phenomenology, light propagation along a media can be
described by a massless particle in a potential. The wave
function obeys the Klein-Gordon equation. One can solve the
stationary state equation and use the eigenstates with positive
energy only. The above argument should remain valid. The
potential should be semi-positive-definite, then the velocity of
photons should not exceed 1 in real time and it will spend
imaginary time on the barrier region.

The key issue is to interpret the meaning of the imaginary time.
From the viewpoint of the intrinsic time of the particle, it
experiences a real time as traveling from the left. After
entering
the barrier it experiences an imaginary time, its behaviour is no
longer deterministic and causal. Even for a hermitian
Hamiltonian, the
evolution along imaginary time becomes nonunitary. In our case,
the wave function decays exponentially as shown by eq. (13).
It takes no real
time to pass the barrier and emits from the right turning point
instantly, and then resumes the lorentzian evolution. The
reason that an outside observer can only sense the
real time lapse, is that all observations and human beings'
consciousness are connected with causal and deterministic
elements of any phenomena. The imaginary time can manifest itself
through some stochastic behaviour.

From the viewpoint of the extrinsic time, the evolution of the
whole system is unitary. Even the probability of a particle
tunneling is less than $1$, the total probability of tunneling
and reflecting remains 1.

We experience imaginary time daily. Maybe all stochastic
phenomena in Nature are the manifestation of imaginary time. It
is well known that if one turns to imaginary time from real
time, then a path integral in quantum mechanics becomes a
partition function in statistical physics. Indeed, imaginary time
has become common sense in black hole physics
and quantum cosmology. The fact that a
tunneling particle decays on barrier region is well understood by
using imaginary time. A photon tunneling through the barrier
instantly regardless of its thickness causes the illusion that
light can travel faster than 1. Locally, a photon always travel
with speed 1 in a vacuum. In the barrier region, it travels along
imaginary time. All these are not only consistent with
relativity, but can also be thought of as a toy version of
traveling from one universe to another, or from one region to
another of the same universe through a wormhole. General
relativity seems to offer the possibility of this kind of rapid
intergalactic travel.

\begin{center}
\vspace*{.25in}
\large
\bf
References
\end{center}
\vspace*{.25in}
\rm
\normalsize

[1]. L.A. MacColl, \it Phys. Rev. \rm \bf 40, \rm 621 (1932).

[2]. E.P. Wigner and L. Eisenbud, \it Phys. Rev. \rm \bf 72, \rm
29 (1947). E.P. Wigner, \it Phys. Rev. \rm \bf 98, \rm 145
(1955).

[3]. A.M. Steinberg, P.G. Kwiat and R.Y. Chiao in \it Proceedings
of the XXVIIIth Rencontre de Moriond \rm ed. by Jean Tran Thanh
Van. Editions Frontieres, Gif-sur-Yvette, France (in press).

[4]. L. Brillouin, \it Wave Propagation and Group Velocity \rm
(Academic, New York, 1960).

[5]. R.Y. Chiao, \it Phys. Rev. \bf A 48, \rm R34 (1993).

[6]. R.Y. Chiao, P.G. Kwiat and A.M. Steinberg, \it Scientific
American, \rm 52, August 1993.

[7]. G.W. Gibbons and S.W. Hawking, \it Euclidean Quantum
Gravity, \rm (World Scientific, Singapore, 1992).

[8]. S.W. Hawking, \it Nucl. Phys. \bf B239, \rm 257 (1984).

[9]. D. Sokolovski and J.N.L. Connor, \it Phys. Rev. \bf A 47,
\rm 4677 (1993).

[10]. A.I. Baz', \it Sov. J. Nucl. Phys. \bf 4, \rm 182 (1967);
\bf 5, \rm 161 (1967). V.F. Rybachenko, \it Sov. J. Nucl. Phys.
\bf 5, \rm 635 (1967).

\end{document}